\documentclass{emulateapj}
\usepackage{amsfonts,amsmath,amstext,graphicx,subfigure,color,setspace}
\usepackage[]{hyperref}
\usepackage[dvipsnames]{xcolor}
\newcommand{\sscribe}[1]{_{\rm{#1}}}
\newcommand{\p}[1]{{#1^\prime}}
\usepackage{color}

\begin{document}
	
	\title{Inverse Compton Scattering Spectra of Gamma-Ray Burst Prompt Emission}
	\author{Yue Zhang\altaffilmark{1,2}}
	\author{Jin-Jun Geng\altaffilmark{1,2,3}}
	\author{Yong-Feng Huang\altaffilmark{1,2}}
	\affil{
		$^1$School of Astronomy and Space Science, Nanjing University, Nanjing 210023, People's Republic of China;\\ hyf@nju.edu.cn, gengjinjun@nju.edu.cn\\
		$^2$Key Laboratory of Modern Astronomy and Astrophysics (Nanjing University), Ministry of Education, People's Republic of China \\
		$^3$Department of Physics, Nanjing University, Nanjing 210093, China
	}

	\begin{abstract}
		
		Although the physical origin of gamma-ray burst (GRB) prompt emission is still controversial, synchrotron radiation from accelerated electrons is a promising mechanism.
		It is believed that electrons are accelerated continuously by ultra-relativistic shocks or magnetic reconnections.
		At the same time, these electrons will be cooled via several processes (mainly adiabatic expansion, synchrotron radiation and inverse Compton scattering (ICS)), which regulate the distribution of electrons.
		An extra high-energy spectrum component is expected to be induced by ICS.
		However, the gamma-gamma annihilation effect may eliminate the high-energy photons and prevent the observers from distinguishing the extra component.
		We perform numerical calculations by taking these effects into account and discuss whether the extra ICS component could be observed.
		By exploring the plausible parameter space for relevant quantities of the GRB ejecta, we present the electron distributions and the corresponding spectra of synchrotron radiation and ICS.
		It is found that the extra component is observable only for ejecta with a rather large bulk Lorentz factor.
		A large Lorentz factor means the adiabatic expansion is the leading process in the electron cooling procedure, which makes the low-energy electron distribution spectrum to be relatively hard.
		Therefore the ICS component is more likely to be detected for GRBs that have a hard low-energy photon spectrum.
		
		\keywords{gamma-ray burst: general --- radiation mechanisms: non-thermal --- relativistic processes --- methods: numerical}
		
	\end{abstract}
	
	\section{Introduction}\label{introduction}
	
	The physical origin of the prompt emission of gamma-ray bursts (GRBs) still remains intriguing.
	The emission process could be synchrotron radiation \citep{1994ApJ...432..181M, 1994MNRAS.269L..41M, 2000ApJ...540..704M}, synchrotron self-Compton (SSC) emission \citep{2000ApJ...544L..17P, 2001A&A...372.1071D, 2008MNRAS.384...33K, 2011A&A...526A.110D, 2012MNRAS.424.3192B}, or photospheric thermal radiation affected by Comptonization \citep{1994MNRAS.270..480T,  2009ApJ...700L.141L, 2010MNRAS.407.1033B}.
	The characteristics of GRB prompt spectra provide certain clues on their origins, and is crucial to solving this fundamental and pressing problem.
	A typical prompt GRB spectrum can be well fit with an expression known as the ``Band-function''
	\citep{1993ApJ...413..281B} in which the photon flux follows a smoothly-joined broken power law, while recent systematic analyses show that two extra components exist in several GRBs \citep{2011ApJ...730..141Z}:  a quasi-thermal component \citep{2010ApJ...709L.172R, 2011ApJ...727L..33G,  2012ApJ...757L..31A,  2013ApJ...770...32G}, and a non-thermal component that can be fit as a power law extending to high energies \citep{ 2003Natur.424..749G,  2009ApJ...706L.138A, 2010ApJ...709L.172R, 2010ApJ...716.1178A, 2011ApJ...729..114A}.
	Although we have phenomenological expressions to describe the GRB spectra, the nature of the prompt radiation hasn't been identified firmly yet.

	Among various mechanisms, synchrotron radiation of electrons accelerated in relativistic shocks has been suggested as a leading candidate radiation mechanism \citep{1994ApJ...432..181M, 1998MNRAS.296..275D}, since there could be sufficient energized electrons and the magnetic field in the emission region could be very strong.
	However, a simple synchrotron mechanism in the so-called ``fast-cooling'' regime predicts a low-energy spectrum as $F_\nu \propto \nu^{-1/2}$ \citep{2000MNRAS.313L...1G}, which is incompatible with the observations of $F_\nu \propto \nu^{0}$ in many GRBs \citep{1993ApJ...413..281B, 2000ApJS..126...19P, 2011A&A...530A..21N,  2013ApJ...764...75G}.
	Several synchrotron mechanisms in detailed scenarios have been put forward to reconcile such a conflict.
	For example, the anisotropic pitch angle distribution could produce a harder spectrum \citep{2000ApJ...540..704M,  2000ApJ...543..722L}.
	\cite{2001A&A...372.1071D} suggested that electrons can cool via inverse Compton scattering (ISC), which has been further developed by a few other authors \citep{2009A&A...498..677B, 2009ApJ...703..675N, 2012MNRAS.424.3192B, 2011A&A...526A.110D}.
	Moreover, \cite{2014NatPh..10..351U} found that the spatial decreasing of magnetic field strength in the jet hardens the prompt spectra, due to the fact that electrons injected in earlier time cool slower than that in the constant magnetic field case.
	The recent numerical calculations by \cite{2018ApJS..234....3G} incorporated the adiabatic, synchrotron and ICS cooling mechanisms in spatially-decaying magnetic fields coherently.
	They found that both the adiabatic/ICS cooling processes and the decaying magnetic fields can result in harder low-energy electron spectra.
	Meanwhile, some recent studies \citep{2017ApJ...846..137O, 2018A&A...613A..16R, 2018A&A...616A.138O, 2018arXiv181006965B} support the idea that some observed GRB prompt spectra originate from synchrotron radiation.
	
	An extra component in addition to the Band component extended to high-energy has been detected in several GRBs \citep{2009ApJ...706L.138A, 2011ApJ...729..114A, 2010ApJ...716.1178A}.
	It has been argued that these high-energy photons arise from ICS of the background Band component by relativistic electrons in the jet \citep{1994ApJ...432..181M}.
	\cite{2009ApJ...703..675N} presents an overall analytic approximate expression for the synchrotron-SSC spectra in the Klein–Nishina (KN) regime.
	Several authors have investigated the electron distributions and prompt spectra altered by SSC in the constant $B'$ case \citep{2009A&A...498..677B, 2011A&A...526A.110D, 2012MNRAS.424.3192B} and the decaying magnetic field case \citep{2018ApJS..234....3G}.
	However, the extra component induced by SSC is not included when calculating the prompt spectra.
	On one hand, photons will gain a lot of energy during the SSC process.
	On the other hand, two high-energy photons will convert to electron/positron pairs via the gamma-gamma annihilation \citep{1967PhRv..155.1404G,  2009A&A...498..677B}.
	It will attenuate the intensity of high-energy SSC spectra and may prevent us from detecting them.
	In this paper, the effect is studied in detail.
	Several numerical calculations on the electron cooling process and prompt spectra are displayed.
	We take the adiabatic expansion, synchrotron radiation and inverse Compton scattering cooling processes into consideration, and assume the magnetic fields around the central engine are spatially dependent.
	The calculations are carried out in plausible parameter space restricted by observations.
	In our calculations, all seed photons are assumed to be produced by synchrotron radiation from shock accelerated electrons, therefore the term ``synchrotron self-Compton'' shares the same meaning with ``inverse Compton scattering'' in this paper unless declared explicitly.
	We focus on the issue that whether the SSC component can be distinguished as an extra component in the observed spectra.
	
	The structure of this paper is as follows.
	In Section \ref{electron_cooling}, a detailed description of electron cooling process is given.
	In Section \ref{both_scenario}, we present our numerical results of electron distribution and photon spectrum in plausible parameter space, and discuss whether the SSC components could be discerned in the spectra.
	We summarize our work and discuss the results in Section \ref{summary_discussion}.
	
	\section{Electron Cooling and Radiation}\label{electron_cooling}
	
	Accelerated electrons in a relativistic jet will lose their energy mainly via three kinds of mechanisms \citep{2009A&A...498..677B}: synchrotron emissions, SSC process and adiabatic cooling.
	Thus, the total cooling rate for electrons reads as
	\begin{equation}
	\dot{\gamma}'\sscribe{e} = \dot{\gamma}'\sscribe{e,syn}  + \dot{\gamma}'\sscribe{e,SSC} + \dot{\gamma}'\sscribe{e,adi},
	\end{equation}
	where $\gamma'\sscribe{e}$ is the Lorentz factor of the electron and $\dot{\gamma}'\sscribe{e}$ is its time derivative.
	The prime ($'$) is used to denote the quantities measured in the co-moving frame of the shocked fluid unless declared explicitly in this paper.
	The cooling rates due to each single mechanism is represented by $\dot{\gamma}'\sscribe{e,syn}$, $ \dot{\gamma}'\sscribe{e,SSC}$ and $\dot{\gamma}'\sscribe{e,adi}$, respectively.
	For an electron moving in the magnetic field with a strength of $B'$, and the emission site at $R$ from the central engine, the cooling rate of each mechanism can be expressed as \citep{1970RvMP...42..237B,  1968PhRv..167.1159J, 2008MNRAS.384.1483F, 2012ApJ...761..147U, 2018ApJS..234....3G}
	
	\begin{equation}\label{cooling_rate_syn}
	\dot{\gamma}'\sscribe{e,syn} = - \frac{\sigma\sscribe{T} \p{B} ^{2} \p{\gamma\sscribe{e}}^{2}}{6 \pi m \sscribe{e} c},
	\end{equation}
	\begin{equation}\label{cooling_rate_adi}
	\dot{\gamma}'\sscribe{e,adi} = - \frac{2}{3} \frac{\gamma\sscribe{e}'}{R} \frac{d R}{dt'},
	\end{equation}
	\begin{eqnarray}\label{cooling_rate_ssc}
	\dot{\gamma}'\sscribe{e, SSC}& =& - \frac{1}{m\sscribe{e} c^{2}}
	\frac{3 \sigma \sscribe{T} c}{4 \p{\gamma\sscribe{e}} ^{2}}
	\int_{\nu'\sscribe{min}}^{\nu'\sscribe{max}}
	\frac{n'(\nu') d\nu'}{\nu'}\nonumber \\
	&\times&\int_{\nu'\sscribe{ic,min}}^{\nu'\sscribe{ic,max}}
	h \nu'\sscribe{ic} d\nu'\sscribe{ic} F(q,g),
	\end{eqnarray}
	where $\sigma\sscribe{T}$ is the Thomson cross-section, $c$ is the light speed, $m\sscribe{e}$ is the electron mass,
	$n'(\nu')$ is the seed photon number density,
	$F(q,g) = 2 q \ln q + (1 + 2 q)(1 - q) + \frac{(4 q g)^{2}}{2(1 + 4 q g)} (1 - q)$,
	$g = \frac{\gamma'\sscribe{e} h \nu'}{m\sscribe{e} c^{2}}$,
	$w = \frac{h \nu'\sscribe{ic}}{\gamma'\sscribe{e} m\sscribe{e} c^{2}}$, and
	$q = \frac{w}{4 g (1 - w)}$.
	The lower and upper limit of the second integration variable $\nu'\sscribe{ic}$ is $\nu'\sscribe{ic,min} = \nu'$ and $\nu'\sscribe{ic,max} = \frac{\gamma'\sscribe{e} m\sscribe{e} c^{2}}{h} \frac{4 g}{4 g + 1}$, respectively.
	For simplicity, the secondary ICS is omitted in our consideration.
	Considering the geometry of the magnetic field in GRB outflows, $B'$ will not remain a constant at different distances from the central engine.
	It has been argued that the transverse (toroidal) component of the magnetic field decays as $B'\sscribe{t} \propto r^{-1}$ while the radial (poloidal) magnetic field component decreases as $B'\sscribe{r} \propto r^{-2}$  \citep{2001A&A...369..694S,  2014NatPh..10..351U}.
	Here we assume that the effective $B'$ evolves with the shell expanding as a power-law form, i.e., $B'(r)  = B'_{0} (r/r_{0})^{-s}$.
	We adopt $s = 1$ throughout this paper, since the toroidal component may play a more significant role.

	The continuity equation of electrons in the energy space is written as
	\begin{equation}\label{continuity_equation}
	\frac{\partial}{\partial t'} \left(\frac{dN'\sscribe{e}}{d\gamma'\sscribe{e}}\right) +
	\frac{\partial}{\partial \gamma'\sscribe{e}}\left[\dot{\gamma}'\sscribe{e,tot} \left(\frac{dN'\sscribe{e}}{d\gamma'\sscribe{e}}\right)\right] =
	Q'(\gamma'\sscribe{e}, t').
	\end{equation}
	Here $dN'\sscribe{e}/d\gamma\sscribe{e}'$ and $Q'(\gamma\sscribe{e}', t')$ are the instantaneous spectrum and the source function of accelerated electrons, respectively.
	These electrons are injected into the emission region at a rate of $N'\sscribe{inj} = \int_{\gamma\sscribe{e,m}'}^{\infty} Q'(\gamma\sscribe{e}', t') d\gamma\sscribe{e}'$.
	We neglect the details of particular electron acceleration mechanisms (e.g., shocks due to internal collisions \citep{2000ApJ...535..152K} or magnetic dissipation events \citep{2011ApJ...726...90Z,  2009MNRAS.395..472K, 2009MNRAS.394L.117N}) to make the result more generic, and  consider a power-law distribution with a slope $p$, i.e., $Q'(\gamma\sscribe{e}', t') = Q'\sscribe{0}  (t') (\gamma\sscribe{e}'/\gamma\sscribe{e,m}')^{-p}$ for $\gamma\sscribe{e}' > \gamma\sscribe{e,m}$.
	Unless declared explicitly, $p = 2.8$ is chosen in our calculations.
	
	The synchrotron radiation power at frequency $\nu'$ in the co-moving frame is given by
	\begin{equation}
	P'\sscribe{syn}(\nu') = \frac{\sqrt{3} q\sscribe{e}^{3} B'}{m\sscribe{e} c^{2}}
	\int_{\gamma\sscribe{e, min}'}^{\gamma\sscribe{e, max}'} \left(\frac{dN'\sscribe{e}}{d\gamma'\sscribe{e}}\right)
	F\left(\frac{\nu'}{\nu'\sscribe{c}}\right) d \gamma'\sscribe{e}.
	\end{equation}
	Here the characteristic frequency $\nu\sscribe{c} = 3 q\sscribe{e} B' \p{\gamma\sscribe{e}}^{2} /(4 \pi m\sscribe{e} c)$, $F(x) = x \int_{x}^{+\infty} K_{5/3}(k) dk$, where $K_{5/3} (k)$ is the modified Bessel function of $5/3$ order.
	Then the SSC seed photon number density $n'$ can be written as \citep{2008MNRAS.384.1483F}
	\begin{equation}
	n'(\nu') \simeq \frac{1}{4 \pi R^{2}} \frac{1}{c h \nu'} P\sscribe{syn}'(\nu').
	\end{equation}
	By integrating over all seed photons and electrons, the SSC power at frequency $\nu'$ is written as \citep{1970RvMP...42..237B, 2008MNRAS.384.1483F}
	\begin{eqnarray}
	P'\sscribe{SSC}(\nu'\sscribe{ic}) &=& \frac{3 \sigma\sscribe{T} c h \nu'\sscribe{ic}}{4}
	\int_{\nu'\sscribe{min}}^{\nu'\sscribe{max}} \frac{n'(\nu') d\nu'}{\nu'}\nonumber\\
	&\times& \int_{\gamma'\sscribe{e,min}}^{\gamma'\sscribe{e,max}} \frac{F(q, g)}{\p{\gamma\sscribe{e}}^{2}}
	\frac{d N'\sscribe{e} (\gamma'\sscribe{e})}{d \gamma'\sscribe{e}} d\gamma'\sscribe{e}.
	\end{eqnarray}
	
	The gamma-gamma annihilation effects will attenuate the intensity of synchrotron and SSC components by
	\begin{equation}
	P'\sscribe{\gamma\gamma}(\nu') = \frac{T'}{4 \pi R^{2}} P'(\nu') \int_{\nu'\sscribe{min}}^{\nu'\sscribe{max}} \frac{ \sigma\sscribe{\gamma\gamma}(\nu', \tilde{\nu}')
		P'(\tilde{\nu}')}{h \tilde{\nu}'}
	d\tilde{\nu}',
	\end{equation}
	where $T' \approx \Delta'/c$ is the time that photons travel through the ejecta, and $\Delta' \approx R/\Gamma$ is the ejecta width in the co-moving frame.
	The cross section of gamma-gamma annihilation is expressed as  \citep{1967PhRv..155.1404G,  2009A&A...498..677B}
	\begin{eqnarray}\label{gamma-gamma-1}
	\frac{\sigma\sscribe{\gamma\gamma} (\nu', \tilde{\nu}')}{\pi r_{0}^{2}/2} &=&
	\left(
	\frac{1 + y^{2}}{1- y^{2}} - y^{2} - \ln \frac{1 + y}{1 - y} + 4 \ \ln\frac{2}{1 - y} \right) \ln\frac{1 + y}{1 - y} \nonumber\\
	&-&\frac{4 y}{1 - y^{2}} + 2 y - \int_{1}^{(1+y)/(1 - y)}
	\frac{\ln (1+x) \ dx}{x},
	\end{eqnarray}
	where
	\begin{equation}\label{gamma-gamma-2}
	y = \sqrt{\frac{h \nu' h \tilde{\nu}' - (m \sscribe{e} c^{2})^{2}}{h \nu' h \tilde{\nu}'}},
	\end{equation}
	and $r_{0} = q\sscribe{e}^{2} / m\sscribe{e} c^{2}$ is the classical electron radius.

	With the competition of these mechanisms, the total radiation power is
	\begin{equation}
	P'(\nu') = P'\sscribe{syn}(\nu') + P'\sscribe{SSC}(\nu') - P'\sscribe{\gamma\gamma}(\nu').
	\end{equation}
	This equation is solved iteratively in our study.
	The observed spectral flux is then
	\begin{equation}
	F(\nu\sscribe{obs}) = \frac{(1 + z) \Gamma P'(\nu'(\nu\sscribe{obs}))}{4 \pi D\sscribe{L}^{2}},
	\end{equation}
	where $\Gamma$ is the bulk Lorentz factor of the ejecta, and $\beta = \sqrt{1 - 1/\Gamma^{2}}$.
	Here $\nu'(\nu\sscribe{obs}) = (1 + z) \Gamma (1 - \beta \cos \theta) \nu \sscribe{obs}$ is the corresponding photon frequency in the co-moving coordination, $D\sscribe{L}$ is the luminosity distance at the cosmological redshift $z$ ($z = 1$ is adopted in our calculations), and $\theta$ is the observation angle.
	We assume that the emission region is at $R_{0} = 10^{15}\ \rm{cm}$.
	This epoch is denoted as $T\sscribe{obs} = 0$.
	
	Since the peak frequency and flux of GRB spectra are determined by the synchrotron mechanism, they could be estimated as
	\begin{equation}\label{nu_peak}
	\nu\sscribe{peak} = \frac{1}{1+z}  \Gamma \gamma'\sscribe{e,a} \frac{3 q\sscribe{e} B'}{4 \pi m\sscribe{e} c},
	\end{equation}
	and
	\begin{equation}\label{flux_peak}
	F\sscribe{\nu}(\nu\sscribe{peak}) = N\sscribe{e} \frac{1+z}{4 \pi D\sscribe{L}^{2}} \frac{\sqrt{3} \Gamma q\sscribe{e}^{3} B'}{m\sscribe{e} c^{2}},
	\end{equation}
	respectively.
	Here $\gamma'\sscribe{e,a}$ is the average electron Lorentz factor, and $N\sscribe{e}$ is the number of relativistic electrons at about $\gamma'\sscribe{e,a}$ in the source.
	$\gamma'\sscribe{e,a}$ could be approximated as $\gamma'\sscribe{e,m}$ in the very beginning of the injection.
	
	The upper and lower integration limits of the electron Lorentz factor, $\gamma'\sscribe{e,min}$ and $\gamma'\sscribe{e,max}$, are set as $10^{1}$ and $10^{7}$ in our calculations, respectively.
	The essential task in our calculations is to solve Equation (\ref{continuity_equation}).
	This work follows the numerical procedures of \cite{2018ApJS..234....3G}, where constrained interpolation profile (CIP) method \citep{1991CoPhC..66..219Y, 2001JCoPh.169..556Y} is utilized to solve the equation effectively and accurately.
	For more details of the numerical methods, please see \cite{2001JCoPh.169..556Y} and Appendix A of \cite{2018ApJS..234....3G}.
	
	\section{Parameters and Results}\label{both_scenario}
	
\begin{figure}
		\centering
		\includegraphics[scale=0.6]{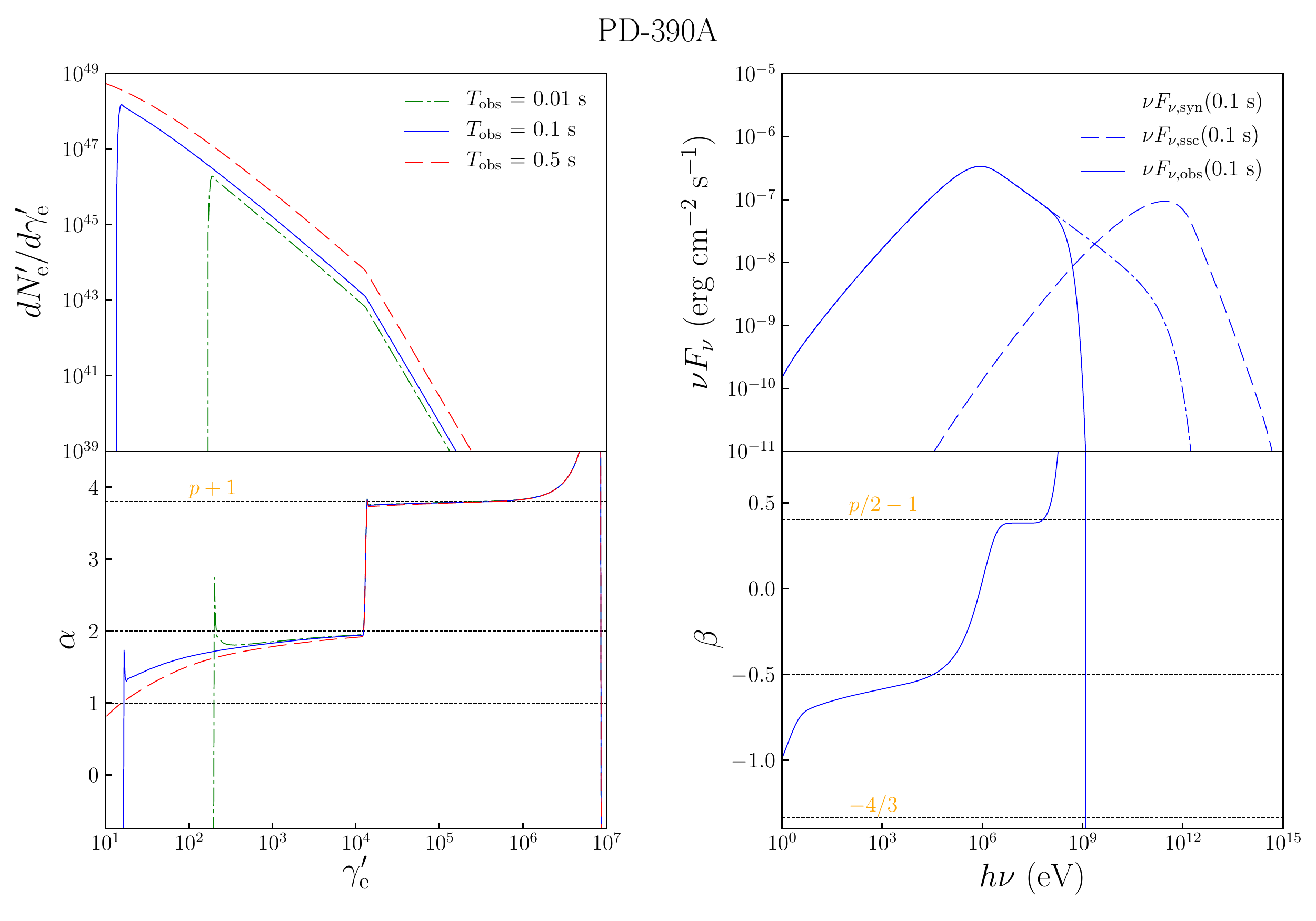}
		\caption{
			Electron distribution and spectra corresponding to PD-390A.
			The evolution of electron distributions is presented in the upper left figure for $T\sscribe{obs} = 0.01\ \rm{s}$ (green dot-dashed), $0.1\ \rm{s}$ (blue solid) and $0.5\ \rm{s}$ (red), respectively.
			The electron spectrum index $\alpha$, defined as $dN'_{\rm{e}}/d\gamma'\sscribe{e} \propto \gamma'^{-\alpha}\sscribe{e}$, is displayed in the lower left figure.
			The upper right figure shows the synchrotron radiation (blue dot-dashed), SSC (blue dashed), and the observed spectra (blue solid) after taking the gamma-gamma annihilation into consideration at $T\sscribe{obs} = 0.1\ \rm{s}$.
			The observed spectrum index $\beta$, defined as $\nu F_{\nu} \propto \nu^{-\beta}$, is displayed in the lower right figure.
		}
		\label{PD-390A}
	\end{figure}

\begin{figure}
		\centering
		\includegraphics[scale=0.6]{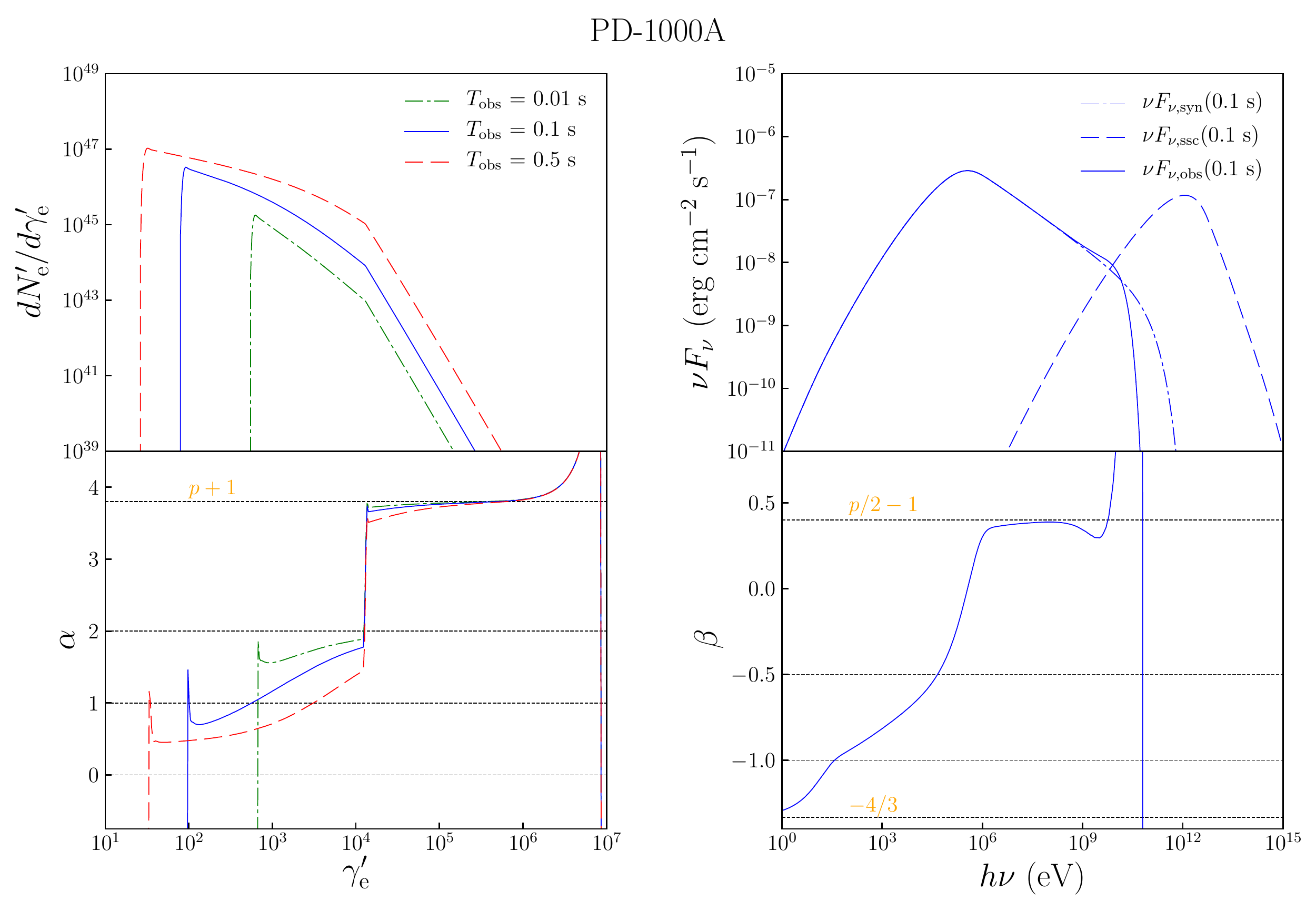}
		\caption{Electron distribution and photon spectra and for PD-1000A.	}
		\label{PD-1000A}
	\end{figure}

\begin{figure}
		\centering
		\includegraphics[scale=0.6]{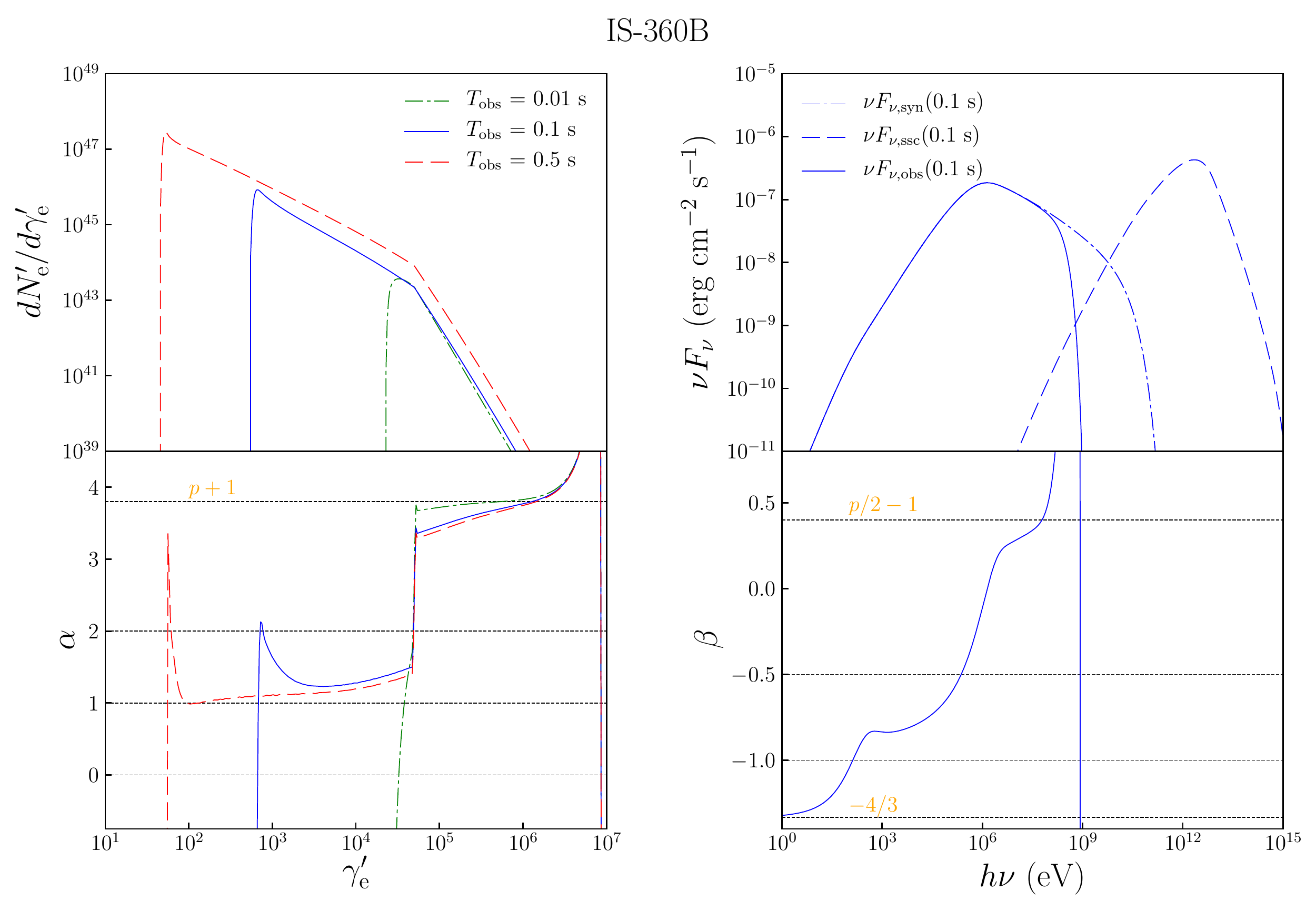}
		\caption{Electron distribution and spectra corresponding to IS-360B.}
		\label{IS-360B}
	\end{figure}

\begin{figure}
		\centering
		\includegraphics[scale=0.6]{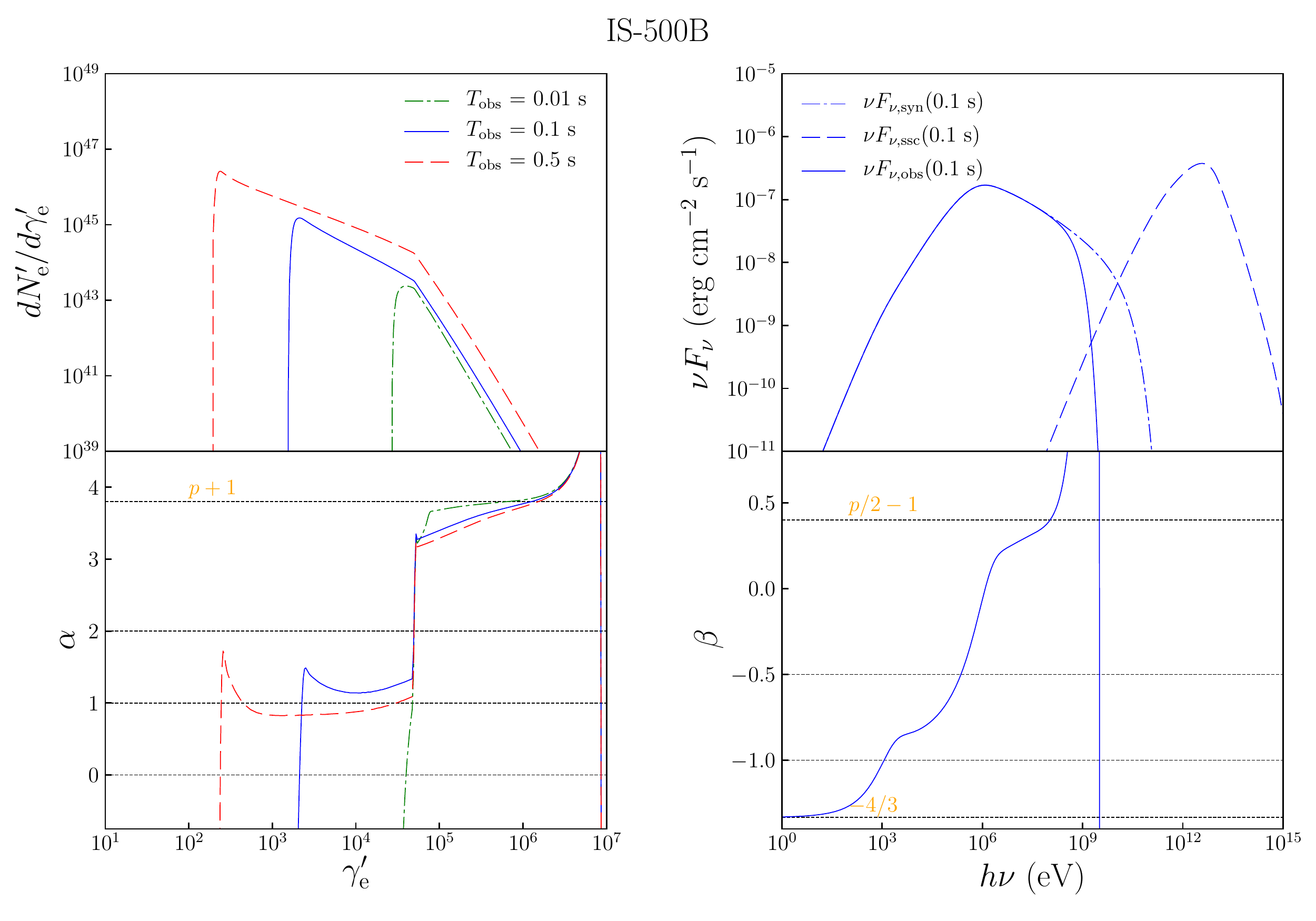}
		\caption{Electron distribution and spectra corresponding to IS-500B.}
		\label{IS-500B}
	\end{figure}

\begin{figure}
		\centering
		\includegraphics[scale=0.6]{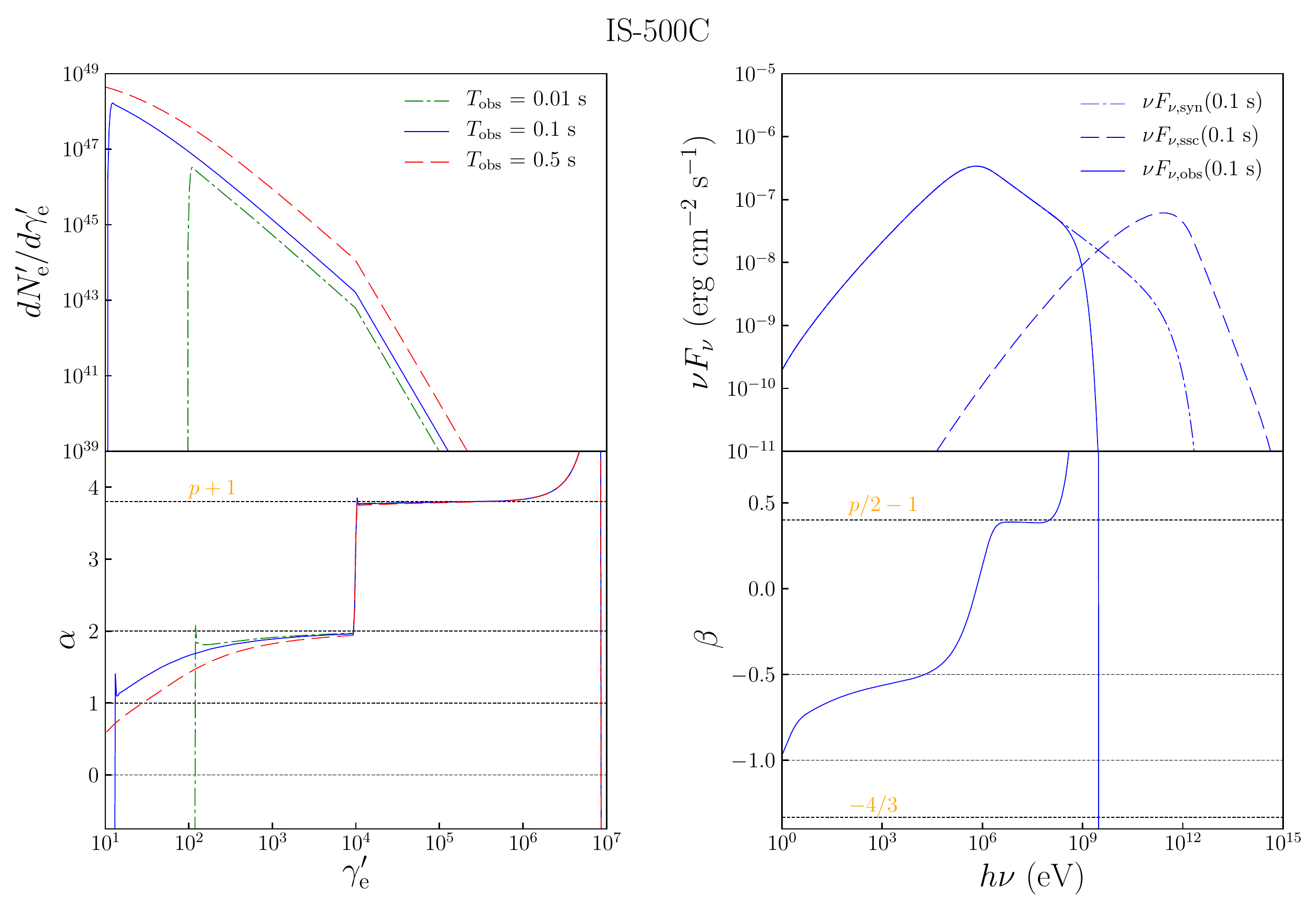}
		\caption{Electron distribution and spectra corresponding to IS-500C.}
		\label{IS-500C}
	\end{figure}

\begin{figure}
		\centering
		\includegraphics[scale=0.6]{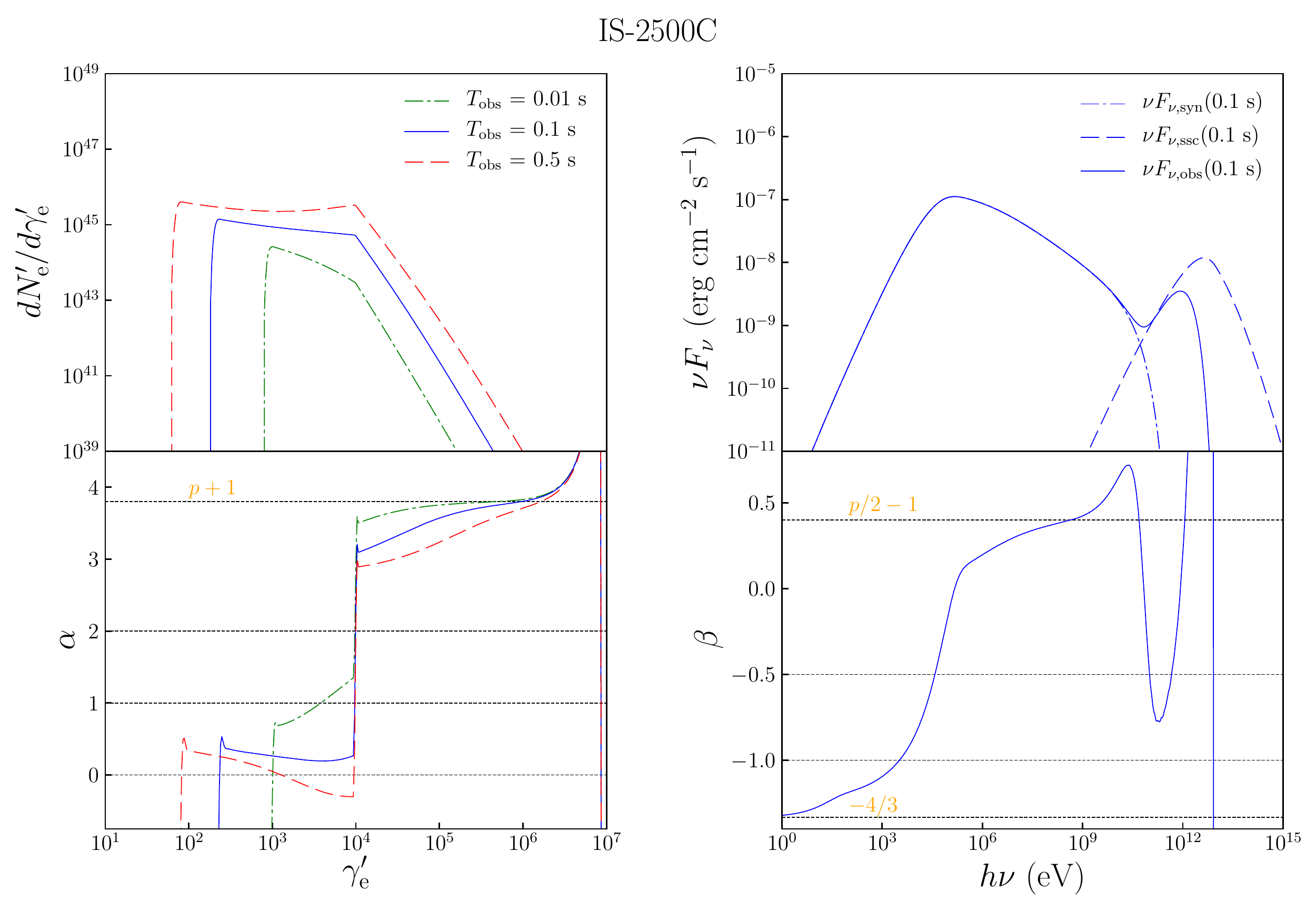}
		\caption{Electron distribution and spectra corresponding to IS-2500C.}
		\label{IS-2500C}
	\end{figure}

\begin{figure}
		\centering
		\includegraphics[scale=0.6]{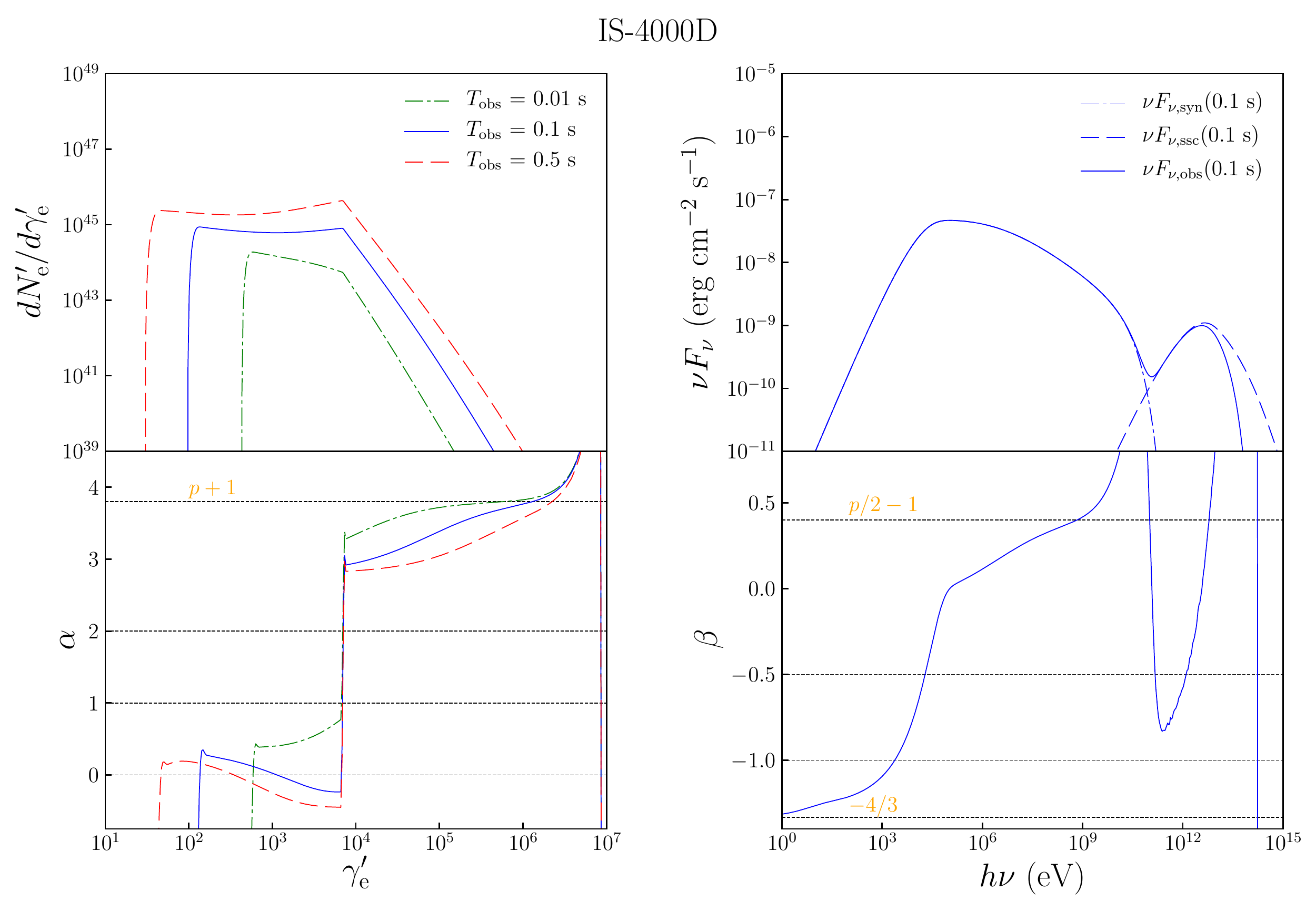}
		\caption{Electron distribution and spectra corresponding to IS-4000D.}
		\label{IS-4000D}
	\end{figure}

	The parameter space of $\Gamma$, $B'_{0}$, $\gamma\sscribe{m}$, and $N\sscribe{inj}'$ for GRB prompt radiation can be constrained from observations \citep{2008MNRAS.384...33K, 2013ApJ...769...69B,  2014MNRAS.445.3892B, 2018ApJS..234....3G}.
	We adopt typical values of GRB peak energy and flux, $E\sscribe{peak} \simeq 500\ \rm{keV}$ and $F\sscribe{peak} \simeq 1\ \rm{mJy}$, then these four parameters are no longer totally independent (see Equations (\ref{nu_peak}) and (\ref{flux_peak})).
	There are two representative scenarios for the outflow ejecta \citep{2004RvMP...76.1143P, 2015PhR...561....1K}, the internal shock \citep{1994ApJ...430L..93R, 1998MNRAS.296..275D, 1999ApJ...522L.105P}, and the Poynting flux-dominated jet \citep{1994ApJ...432..181M, 2002A&A...391.1141D, 2004ASPC..312..449L,  2011ApJ...726...90Z}.
	In the Poynting-flux dominated jet scenario, the isotropic equivalent magnetic luminosity, $L\sscribe{B} \simeq \frac{B'^{2}}{8 \pi} 4 \pi R^{2} c \Gamma^{2}$, is assumed to be larger than the kinetic power of injected electrons, $L\sscribe{e} \simeq N\sscribe{inj}' m\sscribe{e} c^{2} \gamma\sscribe{m}' \Gamma^{2}$ \citep{2018ApJS..234....3G}.
	In the internal shock model, the kinetic energy power carried by protons is dominated in the outflow.
	We denote the average number of protons per electron is denoted by $\eta\sscribe{p}$, and the proton kinetic energy power could be derived as $L\sscribe{p}  \simeq \eta\sscribe{p} N\sscribe{inj}' m\sscribe{e} c^{2} \Gamma^{2}$.
	The parameter space hinted from observations for the two scenarios are different
	(see Sections $3$ and $4$ of \cite{2018ApJS..234....3G} for details).
	Here we explore the plausible parameter space, and exhibited some representative results for both scenarios.
	The parameters used in our calculations are presented in Table \ref{parameters}.
	Each parameter group has its unique code name in the form of ``XX-$\Gamma$Y'', where $\Gamma$ denotes the bulk Lorentz factor of the jet, Y is an identifier distinguishing different subgroups, and ``XX'' could be ``PD'' or ``IS'', standing for ``Poynting-flux dominated'' and ``internal shock'' scenarios,  respectively.
	In Figures \ref{PD-390A} to \ref{IS-4000D}, our numerical results of the electron distributions and photon spectra up to $T\sscribe{obs} = 0.5\ \rm{s}$ for these parameter groups are displayed.

	For the scenario of Poynting-flux dominated jet, we calculate two cases, PD-390A and PD-1000A.
	From the results of PD-390A (Figure \ref{PD-390A}), it is noticed that the electron spectra index below the injection Lorentz factor, $\gamma'\sscribe{e,m}$, is approximately $2$.
	The electron distribution in PD-1000A (Figure \ref{PD-1000A}) is more complicated compared to PD-390A.
	The low-energy electron spectrum index is significantly less than $2$, and even could be basically less than $1$ at the spectrum end (Figure \ref{PD-1000A}).
	The electron spectrum in PD-390A is very close to a fast-cooling distribution, in which the synchrotron cooling overwhelms other two kinds of processes and $B'$ remains constant \citep{2000MNRAS.313L...1G, 2007Ap&SS.309..157D}.
	However, in our calculations electrons are cooling at a decreasing rate with the expanding of the shell, since the cooling rate scales as $\dot{\gamma}'\sscribe{e,syn} \propto B'^{2} \propto R^{-2}$ and $\dot{\gamma}'\sscribe{e,adi} \propto R^{-1}$ \citep{2018ApJS..234....3G}, which can be derived from Equations (\ref{cooling_rate_syn}) and (\ref{cooling_rate_adi}).
	Under such conditions, as proposed by \cite{2014NatPh..10..351U}, the low-energy electron spectrum is harder than that in the standard fast-cooling regime.
	Meanwhile, the ratio between adiabatic and synchrotron cooling rates can be estimated as $\gamma'\sscribe{e,syn}/\gamma'_{\rm{e,adi}}\sim 6.5\ R_{0,15}^{2} (R_{0,15} + \frac{2}{1+z} \Gamma^{2}_{3} T\sscribe{obs,-1})^{-1} B'^{2}_{2} \Gamma_{3}^{-1} \gamma'\sscribe{e,4}$  according to Equations (\ref{cooling_rate_syn}) and (\ref{cooling_rate_adi}).
	We adopt the convention $Q\sscribe{x} = Q/10^{x}$ in cgs units hereafter.
	At the epoch $T\sscribe{obs} = 0.1\ s$, the adiabatic expansion is the major cooling process for electrons with $\gamma'\sscribe{e}$ less than about $500$.
	Therefore in the case of PD-1000, the low-energy electron spectrum are not only hardened by the decreasing synchrotron cooling rate \citep{2014NatPh..10..351U}, but also by the adiabatic cooling itself.
	A hard electron spectrum could then be expected for PD-1000.
	The scenario in case PD-390A, however, is different.
	Because $B'$ is large for PD-390A, the magnetic field is so strong that the electron synchrotron cooling timescale, which scales as $t'\sscribe{syn} \propto B'^{-2}$, is very short.
	Meanwhile $\Gamma$ is relatively small for PD-390A, the increment of the radius and the corresponding decrement of the electron cooling rate is negligible on the electron cooling timescale, and naturally, a nearly fast-cooling electron spectrum is expected.
	It can be clearly seen that the observed flux plummets at a cut-off energy $E\sscribe{cut}$ of $\sim 10^{9}\ \rm{eV}$ and $\sim 10^{11}\ \rm{eV}$ for PD-390A and PD-1000A, respectively.
	Above these frequencies, almost no photons could survive to be observed.
	For both PD-390A and PD-1000A, the SSC flux does not exceed the synchrotron flux at $E\sscribe{cut}$.
	As exhibited in Figures \ref{PD-390A} and \ref{PD-1000A}, the SSC component is completely erased out in the observed spectra and only synchrotron components can be observed.
	
	Our calculations for the internal shock scenario are categorized into three subgroups according to different $\gamma'\sscribe{e,m}$.
	In the SSC dominated cooling systems, there is a small warp just above the low energy end of the electron distribution \citep{2018ApJS..234....3G}.
	Such phenomena can be seen in the cases of IS-360B and IS-500B (see Figures  \ref{IS-360B} and \ref{IS-500B}).
	The electron spectrum index below $\gamma'\sscribe{e, m}$ is approximate a constant except for the warp in the tail for IS-360B and IS-500B.
	The hardening of the low-energy electron spectra originates from not only the decaying magnetic fields \cite{2014NatPh..10..351U} but also the SSC effects \citep{2001A&A...372.1071D,  2009ApJ...703..675N, 2012MNRAS.424.3192B, 2018ApJS..234....3G}.
	Almost all photons are annihilated in the frequency range where the SSC flux exceeds synchrotron for IS-360B and IS-500B.
	Thus we could not expect an extra component to be discerned.
	In the case of IS-500C (Figure \ref{IS-500C}), synchrotron radiation is still the main cooling mechanism, since the magnetic fields are quite strong.
	Electrons are also in the fast-cooling regime in PD-390A, although the shell expands rather fast and the adiabatic cooling becomes comparable to the synchrotron process for low-energy electrons like PD-1000A.
	The SSC component seems not likely to be recognized (Figure \ref{IS-500C}), either.
	For the rest two cases, IS-2500C and IS-4000D,  electrons are cooling down primarily via the adiabatic process, hence the electron spectrum index below $\gamma'\sscribe{e,m}$ is very hard, and could be even less than $0$.
	The cut-off energy $E\sscribe{cut}$ is larger than the energy where SSC flux exceeds the synchrotron component.
	A secondary peak emerges at about $10^{12} \sim 10^{13}\ \rm{eV}$ as shown in Figures \ref{IS-2500C} and \ref{IS-4000D}, respectively.
	Therefore, the high-energy SSC component is likely to be observed for both IS-2500C and IS-4000D.
		
	Now we present an analysis on the electron distribution and the observed spectrum below $E\sscribe{cut}$.
	If electrons are distributed as a power-law function, e.g., $dN/d\gamma'\sscribe{e} \propto \gamma'^{-\alpha}\sscribe{e}$, then the synchrotron radiation from them is also approximately a power-law function, e.g., $\nu F\sscribe{\nu} \propto \nu^{-\beta}$, where $\beta = (\alpha - 1)/2 - 1$ is a valid approximation as long as $\alpha > 1$ \citep{1970RvMP...42..237B}.
	For PD-390A and IS-500C which are in the fast-cooling regime (Figures \ref{PD-390A} and \ref{IS-500C}), the electron spectrum index above $\gamma'\sscribe{e,m}$ is about $p+1$, corresponding to a photon spectrum of $\beta=p/2-1$ above $E\sscribe{peak}$.
	In other cases, the electron above $\gamma'\sscribe{e,m}$ deviates from the power-law distribution more or less.
	The $\beta$ value above $E\sscribe{peak}$ is no longer a constant, but varies slightly around $p/2-1$.
	The $\beta$ value for very low-energy photons is derived as $-4/3$ \citep{1970RvMP...42..237B}.
	How $\beta$ increases from  $-4/3$ to about $p/2-1$ with the growth of photon energy is determined by the low-energy electron distribution.
	In the fast cooling regime, the $\beta$ value has a platform at $-1/2$, since the low-energy electron spectrum index is derived as $\alpha = 2$.
	In the fast-cooling like cases of PD-390A and IS-500C (Figures \ref{PD-390A} and \ref{IS-500C}), the electron spectrum indices are somewhat less than $2$, especially at the low-energy end.
	Hence $\beta$ below $E\sscribe{peak}$ is slightly less than $-1/2$.
	For IS-360B and IS-500B (Figures \ref{IS-360B} and \ref{IS-500B}), the approximation $\alpha \simeq 1.4$ can be applied for low-energy electrons, therefore a moderate slope at about $\beta \simeq -0.8$ can be recognized.
	In the rest three cases, PD-1000A, IS-2500C and IS-4000D (Figures \ref{PD-1000A}, \ref{IS-2500C} and \ref{IS-4000D}), the bulk Lorentz factor $\Gamma$ is so large that the adiabatic cooling makes the electron spectrum deviating from a common power-law distribution significantly.
	The corresponding low-energy photon spectra are relatively hard due to the hard low-energy electron spectra.
	Since the SSC components are not eliminated due to a large $\Gamma$ under our calculations, they are more likely to be recognized in GRBs with a hard low-energy spectrum.
	
\begin{figure}
		\centering
		\includegraphics[scale=0.6]{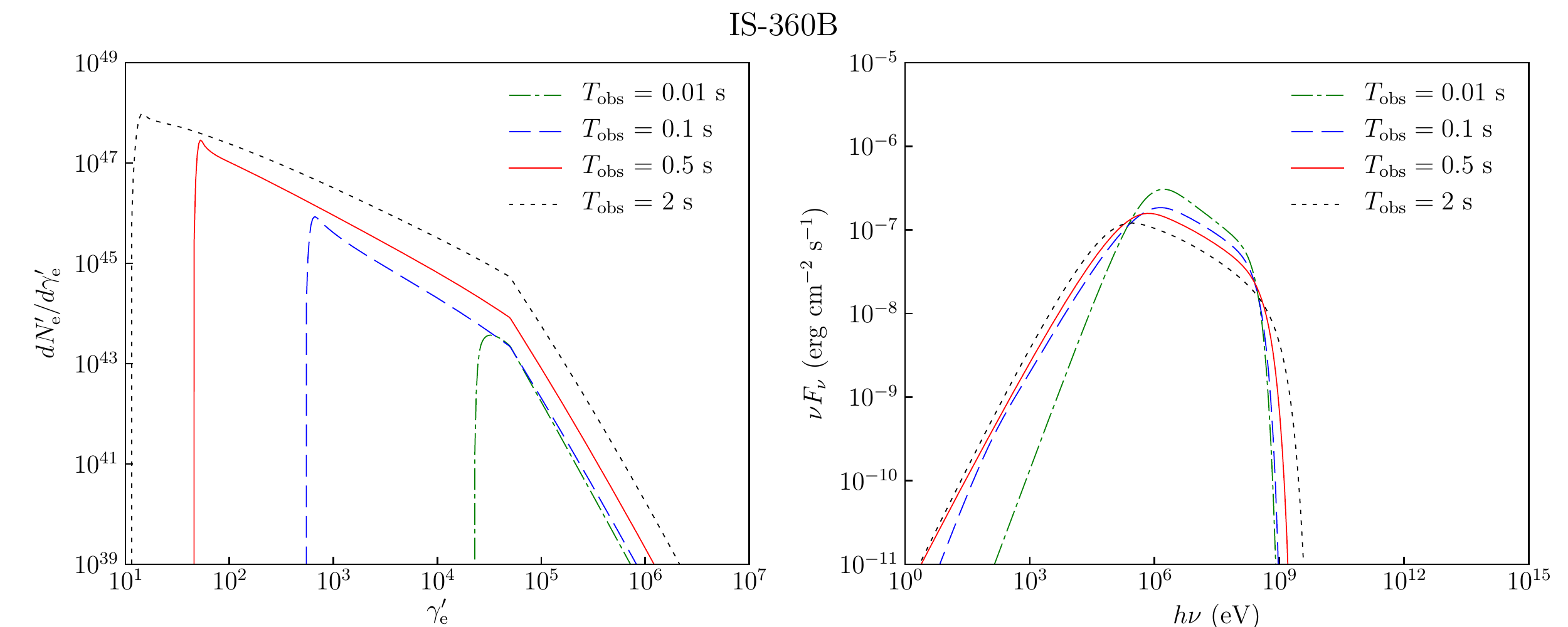}
		\caption{
			The evolution of electron distribution and emission spectra for the case of IS-360B.
			The left panel shows the electron distribution at  
                        $T\sscribe{obs} = 0.01\ \rm{s}$ (dash-dotted line), $0.1\ \rm{s}$ (dashed line), $0.5\ \rm{s}$ (solid line) and $2\ \rm{s}$ (dotted line).
			The evolution of observed spectra is presented in the right panel.
		}
		\label{timeEvolution-IS-360B}
	\end{figure}

\begin{figure}
		\centering
		\includegraphics[scale=0.6]{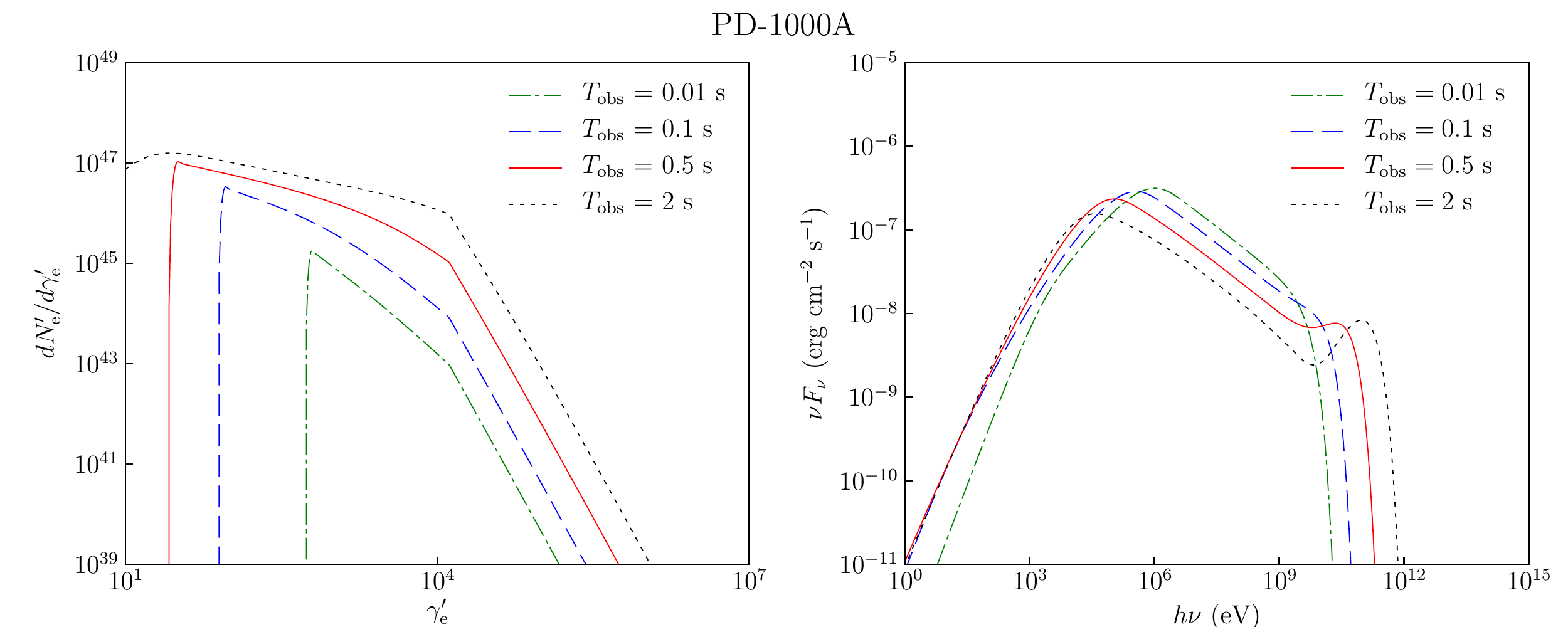}
		\caption{
			The evolution of electron distribution and emission spectra for the case of PD-1000A.
                        The left panel shows the electron distribution at  
                        $T\sscribe{obs} = 0.01\ \rm{s}$ (dash-dotted line), $0.1\ \rm{s}$ (dashed line), $0.5\ \rm{s}$ (solid line) and $2\ \rm{s}$ (dotted line).
			The evolution of observed spectra is presented in the right panel.
		}
		\label{timeEvolution-PD-1000A}
	\end{figure}

\begin{figure}
	\centering
	\includegraphics[scale=0.6]{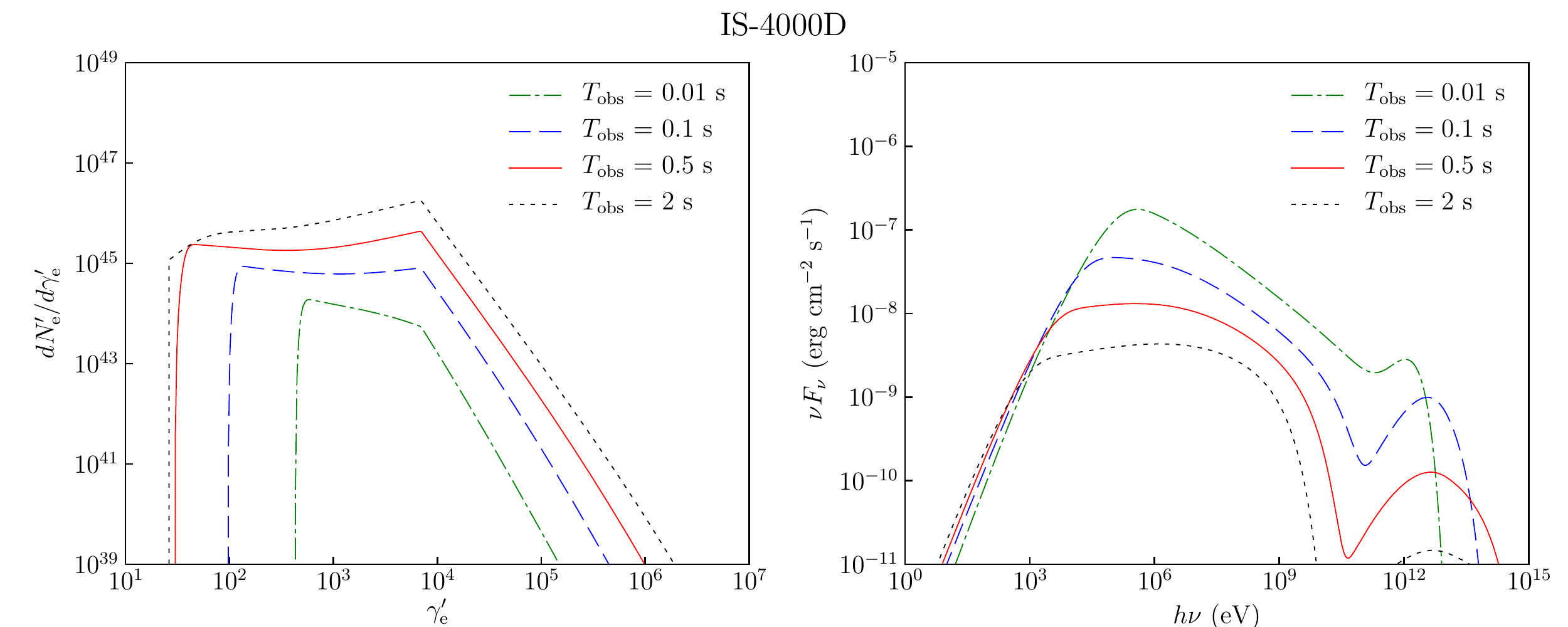}
	\caption{
		The evolution of electron distribution and emission spectra for the case of IS-4000D.
                The left panel shows the electron distribution at  
                        $T\sscribe{obs} = 0.01\ \rm{s}$ (dash-dotted line), $0.1\ \rm{s}$ (dashed line), $0.5\ \rm{s}$ (solid line) and $2\ \rm{s}$ (dotted line).
		The evolution of observed spectra is presented in the right panel.
	}
	\label{timeEvolution-IS-4000D}
	\end{figure}

	In our calculations, the observed spectra are generally calculated for $T\sscribe{obs} = 0.1\ \rm{s}$.
    In reality, the observed GRB spectra usually evolve with time. 
    It is thus necessary to see how our  theoretical spectra evolve during the burst.
    Here let us discuss the spectrum evolution for three parameter groups, IS-360B, PD-1000A and IS-4000D, 
    which has a relatively low, middle and high bulk $\Gamma$, respectively.
    For these three scenarios, we have calculated both the electron distribution and the corresponding observed spectra  
    at $T\sscribe{obs} = 0.01\ \rm{s}$, $0.1\ \rm{s}$, $0.5\ \rm{s}$ and $2\ \rm{s}$.
    The results are shown in Figures \ref{timeEvolution-IS-360B}, \ref{timeEvolution-PD-1000A} and \ref{timeEvolution-IS-4000D}.

    Figures \ref{timeEvolution-IS-360B} shows the spectrum evolution for IS-360B. In this case, since 
    the Lorentz factor is not too large, we see that the general form and the basic characters of the spectra 
    do not change with time significantly. 
    However, a large bulk $\Gamma$ factor usually means a rapid evolution of the electron spectrum.
    In the cases of PD-1000A (Figure \ref{timeEvolution-PD-1000A}) and IS-4000D (Figure \ref{timeEvolution-IS-4000D}), 
    when the Lorentz factor is large enough, we do can observe an obvious evolution in the spectrum. 
    There are two interesting points that should be noted in these two scenarios.
    First, the minimum Lorentz factor of electrons decreases with time, meanwhile the magnetic field strength decays as the shell expands.
    These factors jointly make the peak frequency of the synchrotron component moving toward the low frequency regime. 
    It means that there is a hard-to-soft spectrum evolution. 
    Second, we could see that the electron distribution function is becoming wider and wider with time. 
    As a result, the observed spectrum also expands to a wider and wider frequency range.
    Especially, the cut-off energy at the higher energy end of the spectrum, $E\sscribe{cut}$, increases quickly with time.
    In fact, for PD-1000A and IS-4000D, $E\sscribe{cut}$ at $2\ \rm{s}$ is more than one magnitude larger than that at $0.01\ \rm{s}$.
    This effect may naturally lead to a delay of high energy ($\sim 100$ GeV) photons.

\begin{figure}
	\centering
	\includegraphics[scale=0.6]{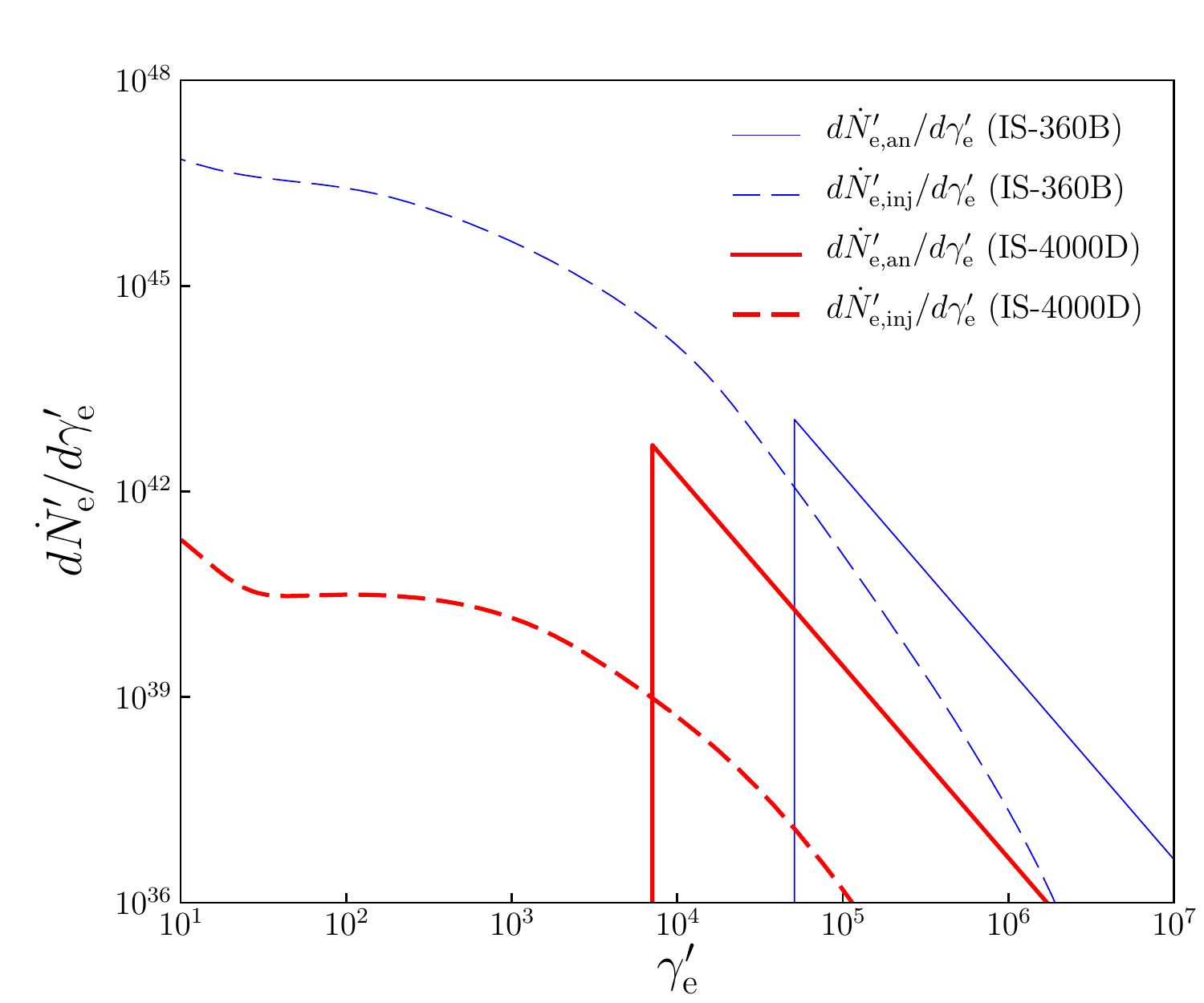}
	\caption{
		The electron production rates due to photon-photon annihilation ($d\dot{N}'_{\rm e, an}/d\gamma'_{\rm e}$, dashed lines),   as compared with the electron  injection rates of GRB acceleration mechanism ($d\dot{N}'_{\rm e, inj}/d\gamma'_{\rm e}$, solid lines).
		The thin lines correspond to IS-360B (with the smallest $\Gamma$ in our parameter set), and the thick lines correspond to IS-4000D  (with the largest $\Gamma$ in our parameter set).
	}
	\label{pair-production}
	\end{figure}

Electron/positron pairs can be produced via photon-photon annihilation. They may be injected into the emission region
and produce additional emission component. Here we give a rough estimation of this pair production rate based on 
the $\delta$-function approximation \citep{1997A&A...325..866B}. 
Basically, photons of energy $\epsilon_{1}$ interact most efficiently with photons of energy 
$\epsilon_{2} = 2/\epsilon_{1}$ \citep{1983Ap.....19..187A}, where the dimensionless photon is defined as 
$\epsilon_i = h \nu_i/m\sscribe{e}c^{2}$. The photon-photon annihilation cross section is  
$\sigma\sscribe{\gamma\gamma} (\epsilon_{1}, \epsilon_{2})  \approx \frac{1}{3}\sigma\sscribe{T} 
\epsilon_{2} \delta\left(\epsilon_{2} - \frac{2}{\epsilon_{1}}\right)$. 
In case of $\epsilon_{1}\gg\epsilon_{2}$ and $\gamma'_{\rm e}\gg1$, the resulting electron pair production rate 
can be estimated as $\frac{d\dot{N}'\sscribe{e,an} (\gamma'\sscribe{e})}{d\gamma'\sscribe{e} }= 
\frac{4 \sigma\sscribe{T} c}{3}  \frac{n\sscribe{ph}(\epsilon_{0}) n\sscribe{ph}(1/\epsilon_{0})}{\epsilon_{0}}$, 
where $n\sscribe{ph}$ is the number density of photons. Assuming that the power of the produced electron/positron 
pairs equals to that of the annihilated photons, i.e., $\int \frac{d\dot{N}'\sscribe{e,an} (\gamma'\sscribe{e})}
{d\gamma'\sscribe{e} } \gamma'\sscribe{e} m\sscribe{e}c^{2} d\gamma'\sscribe{e} \simeq \int \left(P'\sscribe{syn}(\nu') 
+ P'\sscribe{SSC}(\nu') - P'(\nu')\right) d\nu'$, we have calculated the differential pair production rates for the scenarios of 
PD-360B and IS-4000D, which share the smallest and largest bulk Lorentz factor in our parameter groups, respectively.
The results are presented in Figure \ref{pair-production}, where the electron injection rates for both the pair 
production and the normal shock acceleration are displayed. We see that in the higher energy range of 
$\gamma' > \gamma'\sscribe{e,m}$, which plays a dominant role in emission, the pair production rate is 
significantly less than electron injection rate due to normal shock acceleration. 
It means that $\gamma$-ray emission in the prompt GRB phase will not be noticeably affected. 
However, in the lower energy portion of $\gamma' < \gamma'\sscribe{e,m}$, a large number of electrons are 
produced and injected into the emission region. They may give birth to an extra component in the low-energy 
synchrotron spectrum, especially an X-ray bump. This effect is more obvious when the bulk Lorentz facto is 
relatively small. A detailed investigation on this effect is beyond the scope of our current study and could 
be conducted in the future.

	\section{Summary and Discussions}\label{summary_discussion}
	
	Numerical calculations on electron cooling in spatially dependent magnetic fields are performed in this paper, taking the adiabatic expansion, synchrotron radiation and inverse Compton scattering processes into consideration.
	We displayed the computation results of electron distributions and photon spectra for several typical parameter groups in plausible parameter space.
	It is found that the SSC component can be discerned in the observed spectra for parameter groups with a large $\Gamma$.
	The adiabatic process is usually the dominant cooling mechanism for a jet with a large bulk Lorentz factor ($\Gamma>1000$), producing very hard low-energy electron spectra with the spectrum index that could be even less than $0$.
	Therefore the low-energy prompt spectra could be very hard for GRBs with the SSC component detected.

	The SSC spectra peak frequency can be theoretically estimated roughly as follows.
	In the early stage of the radiation, the energy of a significant fraction of radiated photons are around $h\nu'\sscribe{e} \simeq E'\sscribe{peak} = \frac{1+z}{2\Gamma} E\sscribe{peak}$.
	The SSC effect between a typical electron with $\gamma'\sscribe{e} \simeq \gamma'\sscribe{e,m}$ and a typical photon is in the KN limit for most of our parameter groups, since the condition
	$\frac{\gamma'\sscribe{e,m} E'\sscribe{peak}}{m\sscribe{e} c^{2}} \simeq \frac{1+z}{2} \frac{E\sscribe{peak}}{500 \rm{keV}} \frac{\gamma'\sscribe{e,m}}{\Gamma} >1$ is satisfied.
	The typical scattered photon energy in the observer frame can be reckoned as $E\sscribe{SSC} = \frac{2 \Gamma}{1 + z} \gamma'\sscribe{e} m\sscribe{e} c^{2} \simeq 5 \times 10^{12}\times \frac{2}{1+z} \gamma'\sscribe{e, 4} \Gamma_{3}  \ \rm{eV}$ \citep{1970RvMP...42..237B}.
	However, during the cooling procedure, the numbers of low-energy electrons and synchrotron photons increase, and so is the numbers of high-energy electrons and photons.
	The SSC process between these low-energy electrons and photons could be in the Thomson regime, thus the scattered photon energy could be much less than $E\sscribe{SSC}$.
	Besides, SSC scattered photons with energy higher than $E'\sscribe{peak}$ could be produced by the ICS process between electrons and synchrotron photons with energy higher than typical ones.
	Therefore the SSC spectra could be quite wide, as can be seen in Figures \ref{PD-390A} to \ref{IS-4000D}.
	
	According to Equations (\ref{gamma-gamma-1}) and (\ref{gamma-gamma-2}), a photon with an energy $h \nu' > m\sscribe{e} c^{2}$ is apt to be annihilated with another photon.
	Above a particular characteristic frequency, almost all photons are annihilated and a ``cut-off'' in the spectra forms \citep{2012ApJ...753..176L}.
	According to the frequency transformation relation between observer and co-moving frame,
	a large $\Gamma$ indicates a higher observed spectra cut-off.
	Meanwhile, the time that photons travel within the ejecta is shorter for a larger $\Gamma$, so the gamma-gamma annihilation effects are diminished further.
	Thus high energy photons could survive and be observed only for ejecta with a rather large bulk Lorentz factor.
	The detection of photons with an energy of tens $\rm{GeV}$ in a few GRBs leads to the suggestion that $\Gamma \gtrsim 600$ \citep{2009ApJ...706L.138A, 2009Sci...323.1688A}, which is also consistent with our numerical calculations.

    	High energy photons (up to several hundred GeV) have been observed from some GRBs, such as from  
        GRBs 090902B \citep{2009ApJ...706L.138A}, 
        090926A  \citep{2011ApJ...729..114A}, and 130427A \citep{2014Sci...343...42A}. 
        A general feature of such a high energy component is that it usually appears slightly later 
        than the GRB trigger time. For the above three GRBs, the photon energy can be up to tens of GeV and 
        the arrival time of the first GeV photon is delayed by $\sim$ 3 -- 9 s with respect the GRB trigger.  
        In another more interesting case of GRB 090510 \citep{2010ApJ...716.1178A} the observed highest 
        photon energy is about 30 GeV, and the first GeV photon is delayed by $\sim$ 0.7 s.  
        It is interesting to note that such a high energy delay can be naturally explained by the 
        inverse Compton scattering process when the bulk Lorentz factor is large enough. 
        For example, in our scenario of PD-1000A  (Figure \ref{timeEvolution-IS-360B}), the cut-off 
        energy increases quickly over time. As a result, photons up to $\sim 100$ GeV will be observed 
        0.5 -- 2 s after the trigger time. This is roughly consistent with the observed high energy delay 
        in GRBs 090902B, 090926A and 130427A. 
        For an even larger bulk Lorentz factor, such as our case of IS-4000D (Figure \ref{timeEvolution-IS-4000D}), 
        the photons can be up-scattered to even 10 TeV, and again they are expected to be delayed by 0.1 -- 0.5 s. 
        It should also be noted that in GRBs 090902B and 090926A, some high-energy ($\gtrsim 1\ \rm{GeV}$) photons 
        were detected even $\sim 100\ \rm{s}$ after the trigger time. At such a late stage, these GeV photons 
        obviously could not come from the prompt Compton scattering process discussed in this study.  
	They may have a different origin, e.g., being produced by external shocks (i.e. of afterglow origin) or 
        from the proton synchrotron emission \citep{ 2010OAJ.....3..150R}. 
	On such a long timescale, the evolution of the jet bulk Lorentz factor should also be taken into 
        consideration \citep{ 2011MNRAS.415.1663T}.

	In our calculations, we suggest that all seed photons come from the synchrotron emission from accelerated electrons.
	However, various external radiation fields \citep{2016MNRAS.459.3175Y} and photospheric thermal emission \citep{ 1994MNRAS.270..480T, 2000ApJ...529L..17L, 2005MNRAS.361..955B} could also act as the ICS seed photons.
	The Compton upscattering of these photons may result in several additional components superposed on the observed spectra.
	Positrons will emerge in high energy astrophysics phenomena and participate in reactions via pair annihilation and production \citep{1967PhRv..155.1404G}.
	Several authors argued that the electron/positron pair annihilation phenomenon may cause a small peak in the spectrum at around $\sim \Gamma m\sscribe{e} c^{2} /(1+z)$ in the observer coordinate system \citep{ 2004ApJ...613..448P, 2006ApJ...642..995P}.
	We did not include this process in our consideration and omitted the extra components produced by such mechanisms \citep{1982ApJ...258..321S, 2005ApJ...628..857P}.
	The mutual interactions between electrons and other kinds of particles (e.g., protons,  muons) in the plasma jet (e.g., Bethe–Heitler process; see \citep{ 1954PhRv...93..768B, 2013MNRAS.429.3238C}) is also ignored.
	Although the potential components from other kinds of particles are considered to be trivial  \citep{1998ApJ...509L..81T,  1998ApJ...499L.131B, 2007ApJ...671..645A} so that we neglected them in our calculations, a calculation that takes the multiple particle sources and radiation into consideration \citep{2005ApJ...628..857P} may provide a more self-consistent conclusion.
	Besides, when the observer is off-axis, the equal arrival time surface (EATS) effect \citep{1997ApJ...491L..19W,  1998ApJ...493L..31P,  2000MNRAS.316..943H,  2000ApJ...543...90H} should be taken into consideration.
	Further works are needed to improve the results of our current study.

	\section{Acknowledgments}
	We thank the anonymous referee for valuable comments and suggestions.
	This work is partially supported by the National Natural Science Foundation of China (Grants No. 11873030 and 11833003),
	and by the Strategic Priority Research Program of the Chinese Academy of Sciences ``Multi-waveband Gravitational Wave Universe''
        (grant No. XDB23040000).
	JJG acknowledges the support from the National Postdoctoral Program for Innovative Talents (Grant No. BX201700115), the China Postdoctoral Science Foundation funded project (Grant No. 2017M620199).

	\begin{deluxetable}{cccccccc}
		\tabletypesize{\scriptsize}
		\tablewidth{0pt}
		\tablecaption{Parameter Groups Used in the Calculations \label{parameters}}
		\tablehead{%
			\colhead{Parameter Group} &
			\colhead{$\Gamma$} &
			\colhead{$\gamma_{\rm e, m}^{\prime}$} &
			\colhead{$B_0^{\prime}$} &
			\colhead{$N_{\rm inj}^{\prime}$}&
			\colhead{$L_{\rm e}$} &
			\colhead{$L_{B}$} &
			\colhead{$L_{\rm p}$} \\
			\colhead{} &  \colhead{} &  \colhead{($10^4$)}  &  \colhead{($10^2$~G)} &  \colhead{($10^{47}$~s$^{-1}$)} &
			\colhead{($10^{51}$~erg~s$^{-1}$)} &  \colhead{($10^{51}$~erg~s$^{-1}$)} & \colhead{($10^{51}$~erg~s$^{-1}$)}
		}
		\startdata
		PD-390A  &  390   	&  1.3   &  8.8     &   11     & 1.7  & 1.7   					&  - \\
		PD-1000A &  1000   	&  1.3   &  3.4     &   1.6    & 1.7  & 1.7   					&  - \\\hline
		IS-360B  &  360   	&  5 	 &  0.65    &   3.3    & 1.7  & $8.0 \times 10^{-3}$    &  6.4 \\
		IS-500B  &  500   	&  5  	 &  0.46	&  	1.7    & 1.7  & $8.0 \times 10^{-3}$    &  6.4 \\
		IS-500C  &  500   	&  1  	 &  11		&  	8.4    & 1.7  & 5  						&  32 \\
		IS-2500C &  2500   	&  1   	 &  2.3  	& 	0.34   & 1.7  & 5  						&  32 \\
		IS-4000D &  4000   	&  0.7	 &  2.9 	& 	0.19   & 1.7  & 21  					&  46
		\enddata
	\end{deluxetable}
	
\end{document}